\documentclass[aps,prl,twocolumn,showpacs,superscriptaddress,floatfix]{revtex4-2}

\usepackage{graphicx}% Include figure files
\usepackage{dcolumn}% Align table columns on decimal point
\usepackage{bm}% bold math
\usepackage{float}
\usepackage{amssymb}
\usepackage{hyperref}
\usepackage{xcolor}
\hypersetup{colorlinks=true,citecolor={blue},linkcolor={blue},urlcolor={blue}}

\begin{document}

\preprint{APS/123-QED}

%\title{Observation of vortex lattice in trapped polariton condensate with beating eigenstates}
\title{Spontaneous formation of time-periodic vortex cluster in nonlinear fluids of light}
% Force line breaks with \\

\author{Kirill A. Sitnik}
\affiliation{Hybrid Photonics Laboratory, Skolkovo Institute of Science and Technology, Territory of Innovation Center Skolkovo, Bolshoy Boulevard 30, building 1, 121205 Moscow, Russia}

\author{Sergey Alyatkin}
\affiliation{Hybrid Photonics Laboratory, Skolkovo Institute of Science and Technology, Territory of Innovation Center Skolkovo, Bolshoy Boulevard 30, building 1, 121205 Moscow, Russia}

\author{Julian D.  T\"opfer}
\affiliation{Hybrid Photonics Laboratory, Skolkovo Institute of Science and Technology, Territory of Innovation Center Skolkovo, Bolshoy Boulevard 30, building 1, 121205 Moscow, Russia}

\author{Ivan Gnusov}
\affiliation{Hybrid Photonics Laboratory, Skolkovo Institute of Science and Technology, Territory of Innovation Center Skolkovo, Bolshoy Boulevard 30, building 1, 121205 Moscow, Russia}

\author{Tamsin Cookson}
\affiliation{Hybrid Photonics Laboratory, Skolkovo Institute of Science and Technology, Territory of Innovation Center Skolkovo, Bolshoy Boulevard 30, building 1, 121205 Moscow, Russia}

\author{Helgi Sigurdsson}
\affiliation{Science Institute, University of Iceland, Dunhagi 3, IS-107, Reykjavik, Iceland}
\affiliation{Department of Physics and Astronomy, University of Southampton, Southampton SO17 1BJ, United Kingdom}

\author{Pavlos G. Lagoudakis}
\email[P.G.Lagoudakis ]{P.Lagoudakis@skoltech.ru}
\affiliation{Hybrid Photonics Laboratory, Skolkovo Institute of Science and Technology, Territory of Innovation Center Skolkovo, Bolshoy Boulevard 30, building 1, 121205 Moscow, Russia}
\affiliation{Department of Physics and Astronomy, University of Southampton, Southampton SO17 1BJ, United Kingdom}

\date{\today}

\begin{abstract}
We demonstrate spontaneous formation of a nonlinear vortex cluster state in a microcavity exciton-polariton condensate with time-periodic sign flipping of its topological charges at the GHz scale. When optically pumped with a ring-shaped nonresonant laser, the trapped condensate experiences intricate high-order mode competition and fractures into two distinct trap levels. The resulting mode interference leads to robust condensate density beatings with periodic appearance of orderly arranged phase singularities. Our work opens new perspectives on creating structured free-evolving light, and singular optics in the strong light-matter coupling regime.
\end{abstract}

\maketitle

Optical vortices~\cite{Coullet1989}, are phase-singular solutions of quantized optical angular momentum, also known as a topological charge, in paraxial electromagnetic fields. Their topological charge defines integer values of the accumulated phase going around the vortex core, characterized by vanishing density of the field. Investigation of vortices and their realization in different optical systems has led to the establishment of the research field of \emph{singular optics}~\cite{DENNIS2009293, Shen_LMAppl2019}. This field has quickly developed from early realizations of optical vortices~\cite{Allen1992}, complex states with fractional vorticity~\cite{Berry2004}, to modern applications~\cite{Qiu_Science2017}. It includes optical tweezers~\cite{Berry2004}, particle manipulation in 3D~\cite{Paterson2001, MacDonald2002}, DNA structure modification~\cite{Zhuang2004}, optical microscopy overcoming the Rayleigh resolution limit~\cite{Tamburini2006}, entanglement based quantum cryptography~\cite{Fickler2016}, and multiplexed communication systems~\cite{Bozinovic2013}.

 Lattices of optical vortices (or so called \emph{vortex crystals}) are elusive and complex photonic states which have been realized in solid-state lasers~\cite{Chu2012, Shen2018}, diode lasers and VCSELs~\cite{Brambilla1991, Scheuer1999}. Unlike conventional vortex lattices in turbulent electron fluids~\cite{Fine_PRL1995}, rotating equilibrium superfluids~\cite{donnelly1991quantized} or Bose-Einstein condensates (BECs)~\cite{pitaevskii2016bose, Fetter_RMP2009}, optical vortex crystals are stabilized by an intricate balance of transverse modes which synchronize over the laser gain. Being a weakly interacting photonic system, they are ill placed to study hard-to-reach physics compared to rotating atomic condensates, such as bosonic fractional quantum Hall effects~\cite{Regnault_PRL2003, Schweikhard_PRL2004}. 

A system which lies at the boundary between the photonic systems and atomic BECs are exciton-polariton condensates. Exciton-polaritons (from here on polaritons) are composite bosons appearing in the strong-coupling regime in semiconductor microcavities that can undergo a power-driven phase transition into a macroscopic coherent state known as a polariton condensate~\cite{Carusotto_RMP2013}. Their strong interactions and small effective mass ($\sim 10^{-5}$ free electron mass) makes them an exciting testbed to study superfluidity~\cite{Amo_NatPhys2009, Amo_Nature2009, Amo_Science2011, Lerario_NatPhys2017} and vorticity~\cite{Lagoudakis2008, Sanvitto_NatPhys2010, Roumpos_NatPhys2011, Donati_PNAS2016, Dominici_NatComm2018, Caputo_NatPhot2019, Kwon_PRL2019, Ma2020, Cookson2021} far from equilibrium. Moreover, the two-component $\sigma^{\pm}$ pseudospin structure of polaritons opens pathways to investigate polarization sensitive vorticity~\cite{Lagoudakis_Science2009, Dominici_SciAdv2015}, vector beams and optical skyrmionic textures~\cite{Donati_PNAS2016, Cilibrizzi_PRB2016} in the strong-coupling regime. Theoretical proposals suggested that complex structured resonant lasers could be used to imprint a polariton condensate vortex lattice~\cite{Liew_PRL2008,
Gorbach_PRL2010} which was later experimentally confirmed~\cite{Boulier2016, Panico2021}. However, a spontaneously self-arranging vortex lattice under nonresonant excitation~\cite{Keeling_PRL2008}, a signature of U(1) symmetry breaking and free-evolving polariton dynamics, has only been reported in the weak tails of multiple spatially overlapping and interfering condensates~\cite{Tosi2012}, and thus, remains elusive.

A well-known method to generate polariton condensates that can display exotic (large nonlinearity) and high-order states is through optical trapping where annular shaped nonresonant excitation beams create complex-valued potentials which provide gain and confinement to polaritons~\cite{Askitopoulos2013, Cristofolini_PRL2013}. The advantage of this method is that the condensate can undergo stimulated scattering to populate higher-order modes~\cite{Askitopoulos_PRB2015, Sun_2018} instead of just the ground state, including vortices~\cite{Dall_PRL2014,PhysRevLett.120.065301, Ma2020}, and even become fractured across multiple modes~\cite{Lotka2020} with consequent nonstationary density self-oscillations.

In this Letter, we demonstrate spontaneous formation of a polariton condensate vortex cluster undergoing cyclic dynamical evolution (indicating a limit cycle) in an optically imprinted trap. The nonresonantly excited condensate state corresponds to a nonstationary superposition of even Ince-Gaussian modes $\Psi = IG_{31}^{(e)}e^{i \Delta t/2\hbar} + IG_{33}^{(e)} e^{-i \Delta t/2\hbar} e^{i \phi}$~\cite{Bandres2004, bentley2006generation} with weak mode splitting $\Delta$ due to slight ellipticity in the trap. The condensate dynamics is characterized by periodic disappearance and reappearance of the vortex cluster with flipped signs of topological charges at the GHz scale. This is confirmed with time-delay interferometric measurements, from which we extract the time-resolved spatial first-order coherence function $g^{(1)}(\mathbf{r},\tau)$ as a function of time delay $\tau$. We observe coherence oscillations of period $T \approx 200$ ps, indicating robustness in the dynamical oscillations and corroborating the presence of a limit cycle behaviour. Moreover, by means of a homodyne interferometric technique~\cite{Alyatkin2020}, where a weak resonant ``seed" laser locks the condensate phase with respect to reference wave, we directly reconstruct the energy resolved condensate spatial intensity and phase maps, corresponding to the split $IG_{31}^{(e)}$ and $IG_{33}^{(e)}$ modes. Unlike previous works~\cite{Tosi2012}, our nonresonant optical approach realises vortices in the dense center of the driven polariton fluid. Our techniques do not imprint any phase or vorticity onto the condensate which rather spontaneously forms and self-organizes, and no trap rotation is implemented to generate a vortex lattice like in superfluids~\cite{donnelly1991quantized} or atomic condensates~\cite{Raman_PRL2001, Schweikhard_PRL2004}.
\begin{figure}
\includegraphics[width=0.97\linewidth]{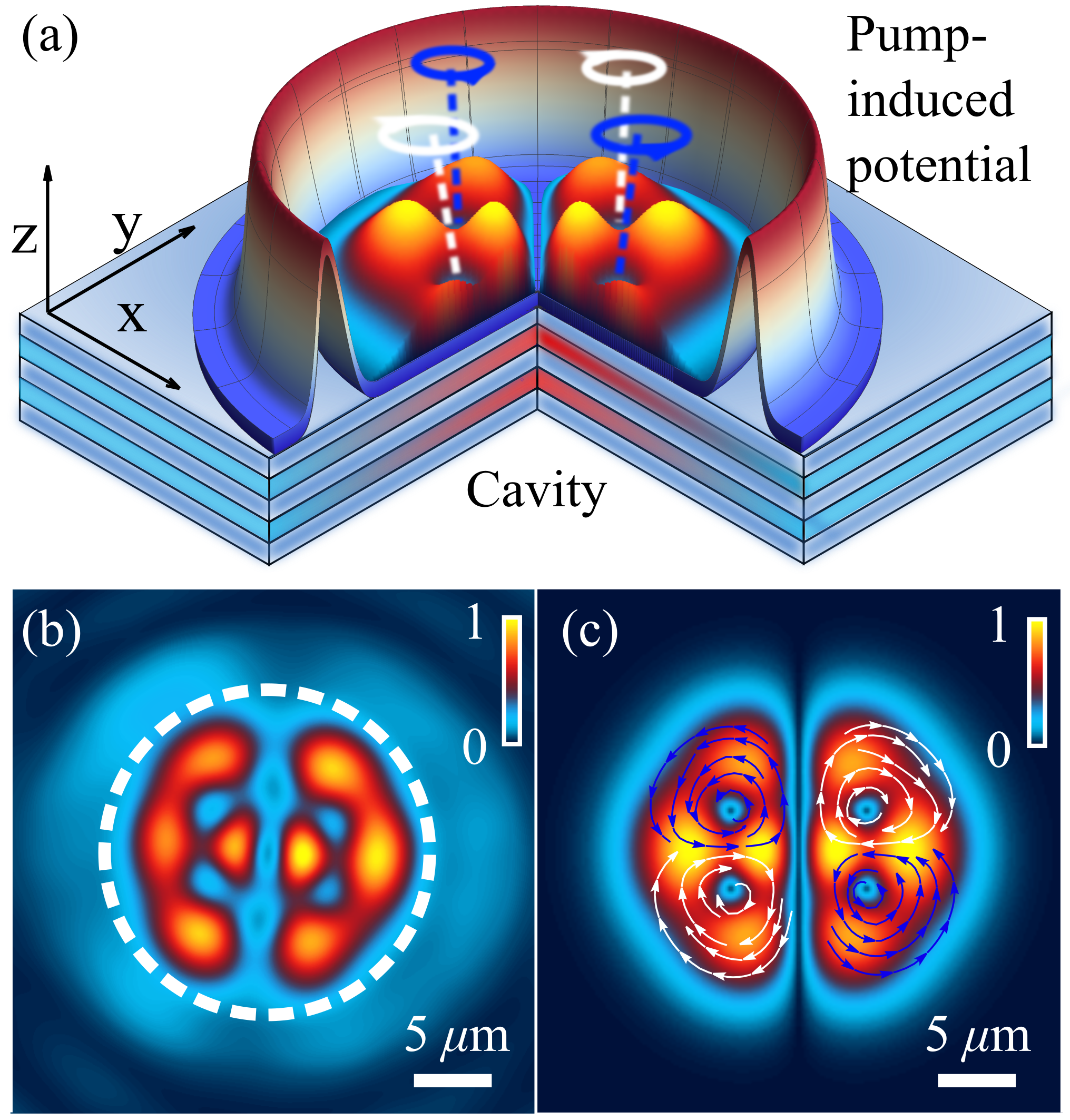}% Here is how to import EPS art
\caption{\label{fig:epsart} (a) Schematic image of the driven polariton condensate (blue-orange surface) trapped in a nonresonantly pump induced annular potential (blue-red surface). (b) Normalized time-integrated measured real-space polariton photoluminescence at pump power of 1.2$\times P_{thr}$, white dashed circle indicates the position of the excitation ring. (c) Normalized corresponding wave function density of two Ince-Gaussian modes $ \langle|\Psi|^2\rangle  =  \langle |IG_{31}^{(e)}e^{ i \Delta t/2\hbar} + IG_{33}^{(e)} e^{-i \Delta t/2\hbar} e^{i \phi}|^2 \rangle$ where $\langle . \rangle$ denotes $t \to \infty$ time-average. White and blue arrows denote the probability currents for the instance $IG_{31}^{(e)} + iIG_{33}^{(e)}$ indicating the vortices positions.}
\label{fig1}
\end{figure}
%\section{Results and discussion}

We use a 2$\lambda$ GaAs-based microcavity with three pairs of embedded InGaAs quantum wells~\cite{doi:10.1063/1.4901814}, excited nonresonantly at the energy of 1.5578 eV with a single-mode continuous-wave laser chopped (at 5 kHz and 1$\%$ duty cycle in all time-integrated measurements) with an acousto-optical modulator to avoid heating. The microcavity is cooled down to a temperature of $\approx$4 K in a closed-cycle cryostat. In order to create an excitation laser profile with ring-shaped intensity distribution, as shown schematically in Fig.~\ref{fig1}(a), we use a programmable phase-only spatial light modulator. At pump power $P=1.2P_{thr}$, where $P_{thr}=44$ mW is the condensation threshold power, we collect the near-field polariton photoluminescence (PL) using a charge-coupled-device camera and observe the trapped polariton condensate with density structure [see Fig.~\ref{fig1}(b)] belonging to higher-order modes of the trap. We note that the imprinted optical trap inherits finite geometric ellipticity from the pump profile. We find good agreement by theoretically constructing the polariton PL in Fig.~\ref{fig1}(c) using a linear superposition of two even Ince-Gaussian modes $\langle |\Psi|^2 \rangle = \langle |IG_{31}^{(e)}e^{i \Delta t/2\hbar} + IG_{33}^{(e)} e^{-i \Delta t/2\hbar} e^{i \phi}|^2 \rangle = |IG_{31}^{(e)} \pm iIG_{33}^{(e)}|^2$. Here $\langle . \rangle$ denotes $t \to \infty$ time-average. The Ince-Gaussians are the natural solutions in systems separable with elliptical coordinates~\cite{Bandres2004}. Overlaid colored arrows in Fig.~\ref{fig1}(c) indicate the polariton current $\mathbf{j} = \text{Re}[-i\Psi^* \nabla \Psi]$ for the example instance $\Psi = IG_{31}^{(e)} + iIG_{33}^{(e)}$ which reveals vortices of topological charge $\pm 1$ (blue and white arrows, respectively). We note that such elliptical solutions were already reported for polaritons in the linear regime in etched microcavity mesas~\cite{Nardin2010}.

We next extract the polariton condensate density through a homodyne interferometric technique~\cite{Alyatkin2020}. We use a Mach-Zehnder interferometer, where the cavity emission is collected in transmission geometry and interfered with a resonant plane reference wave. The reference wave source is a single-mode external cavity diode laser (the linewidth $\approx$ 100 kHz), tuned to the energy of the condensate, which is additionally locally seeded by a focused weak resonant beam (FWHM $\approx$ 2 $\mu$m). It is worth noting that both, the reference wave and the weak seed beam originate from the same laser and, therefore, are phase-locked sources. The energy of the reference wave can be precisely controlled with a wavemeter with an accuracy of $\approx$ 2.5 $\mu$eV, smaller than the condensate linewidth. The reported typical PL linewidth for the trapped condensate is $\leq$ 25 $\mu$eV \cite{cohrev}. We point out that this number approximately coincides with a spectral resolution of our spectrometer and therefore, limits the direct linewidth measurements and an observation of any fine energy splitting.
\begin{figure}[t]
\includegraphics[width=0.97\linewidth]{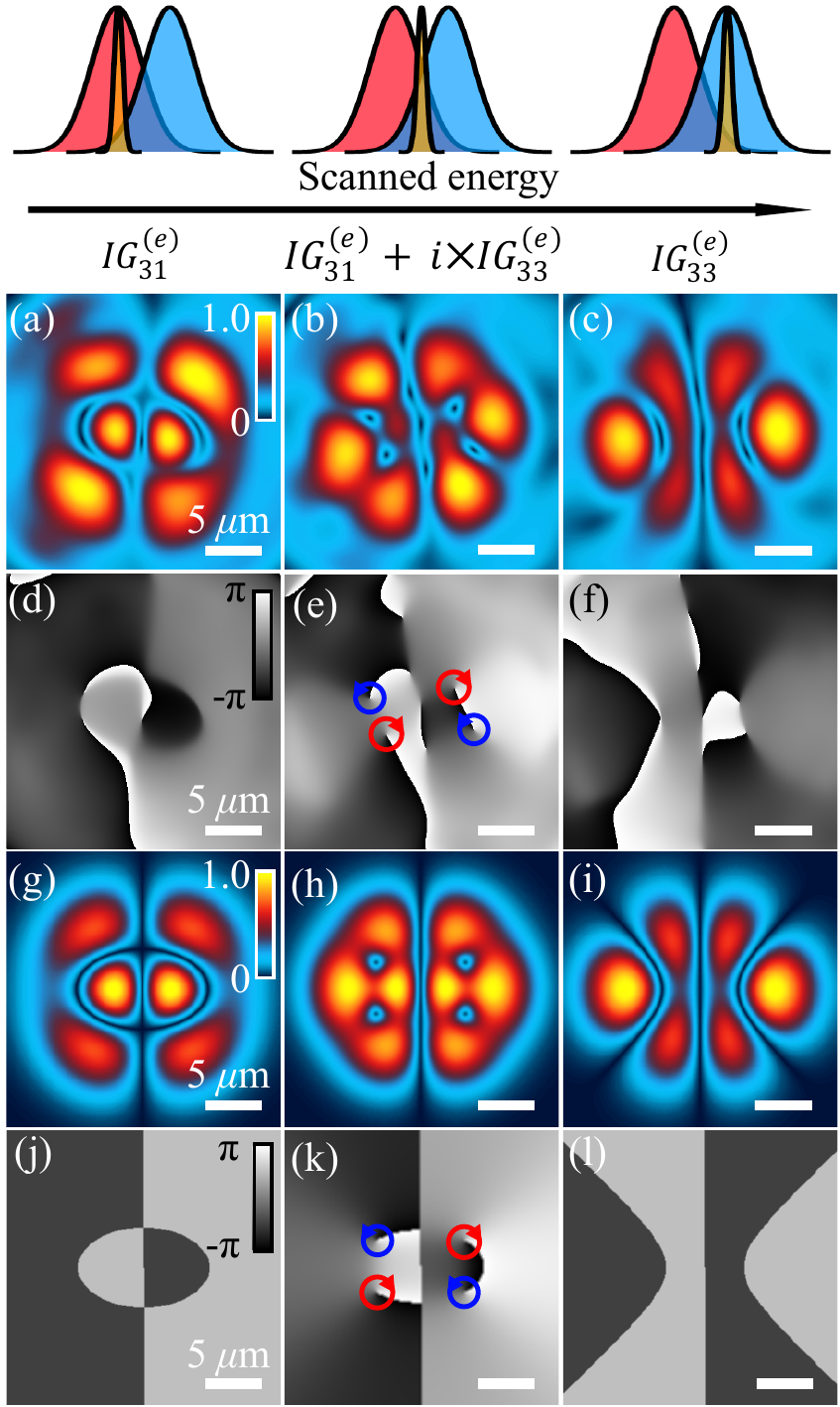}% Here is how to import EPS art
\caption{\label{fig:epsart}(a,b,c) Normalized experimentally measured real-space polariton condensate PL, and (d,e,f) corresponding real-space phase maps, extracted at different energies. (g,i) Normalized spatial intensity distributions of absolute value for the complex functions described by even Ince-Gaussian modes $IG_{31}^{(e)}$ and $IG_{33}^{(e)}$ with an ellipticity parameter $\epsilon$=2, and (j,l) corresponding arguments (phase maps) of the functions. (h) Normalized density and (k) argument for a superposition of the Ince-Gaussians $IG_{31}^{(e)}+iIG_{33}^{(e)}$. Scale bar in all panels corresponds to 5 $\mu$m. All phase distributions (d-f, j-l) plotted for the phase range from $-\pi$ to $\pi$.}
\label{fig2}
\end{figure}

The experimental results on the extracted condensate density and phase are shown in Figs.~\ref{fig2}(a-c) and~~\ref{fig2}(d-f), respectively, where the reference laser was scanned in energy by $\approx$ 25.5$\pm$7.5 $\mu$eV. In contrast to time-integrated results, these images are extracted from corresponding interference (with a plane reference wave) patterns recorded in a ``single-shot" excitation regime. In this regime the polariton system is excited only with an individual pulse with increased pulse width to 50 $\mu$s (compared to 2 $\mu$s pulses arriving every 200 $\mu$s in time-integrated measurements) to detect enough photocounts but avoid detrimental sample heating. Therefore, each recorded ``single-shot" interferogram corresponds to a single realization of the condensate. Next, by means of off-axis digital holography~\cite{Kreis1986, Liebling2004} we extract both the energy-resolved polariton phase and density maps. Figure~\ref{fig2}(a-c) shows several distinct condensate states which can be theoretically reconstructed using Ince-Gaussian modes, as shown in Figs.~\ref{fig2}(g-i), with good agreement. The results evidence that the individual $IG_{31}^{(e)}$ and $IG_{33}^{(e)}$ modes are split in energy and become picked up by the homodyne technique when the seed laser is resonant with each mode. We note that minor differences between theory and experiment can be explained by the weak presence of higher-order states.
When the energy of the seed laser is tuned to be exactly in between the modes [see Figs.~\ref{fig2}(b,e)] the visibility contrast of the corresponding interferogram lowers (see Supplemental Material) and only a small portion of the condensate mode populations that overlap with the seed laser frequency become partially synchronized $\pi/2$ out of phase~\cite{Cerna_2009,Chestnov_NJP2019}. The corresponding extracted phase map in Fig.~\ref{fig2}(e) reveals an arrangement of phase singularities in qualitative agreement with our modelling in Fig.~\ref{fig2}(k), evidencing a condensate vortex cluster state matching a specific superposition of $\Psi = IG_{31}^{(e)} + iIG_{33}^{(e)}$. The reason the condensate picks $\pi/2$ phase difference is attributed to various hard-to-avoid symmetry breaking effects such as sample disorder, pump inhomogeneity, and/or different overlap of the seed beam profile with the Ince-Gaussian modes. The effect of the resonant stimulation is well-known from earlier work~\cite{Cerna_2009}, where coherent control of the wave function of trapped polaritons has been studied. We note that typical seed laser power used in our experiments was limited by 0.016$\%$ of $P_{thr}$ value for nonresonant excitation. Moreover, we emphasize that across all scanned energies the time-integrated condensate PL remained qualitatively the same to the one presented in Fig.~\ref{fig1}(b), underlining that the system always populates the two dominant modes $IG_{31}^{(e)}$ and $IG_{33}^{(e)}$.

\begin{figure}[t!]
\includegraphics[width=0.97\linewidth]{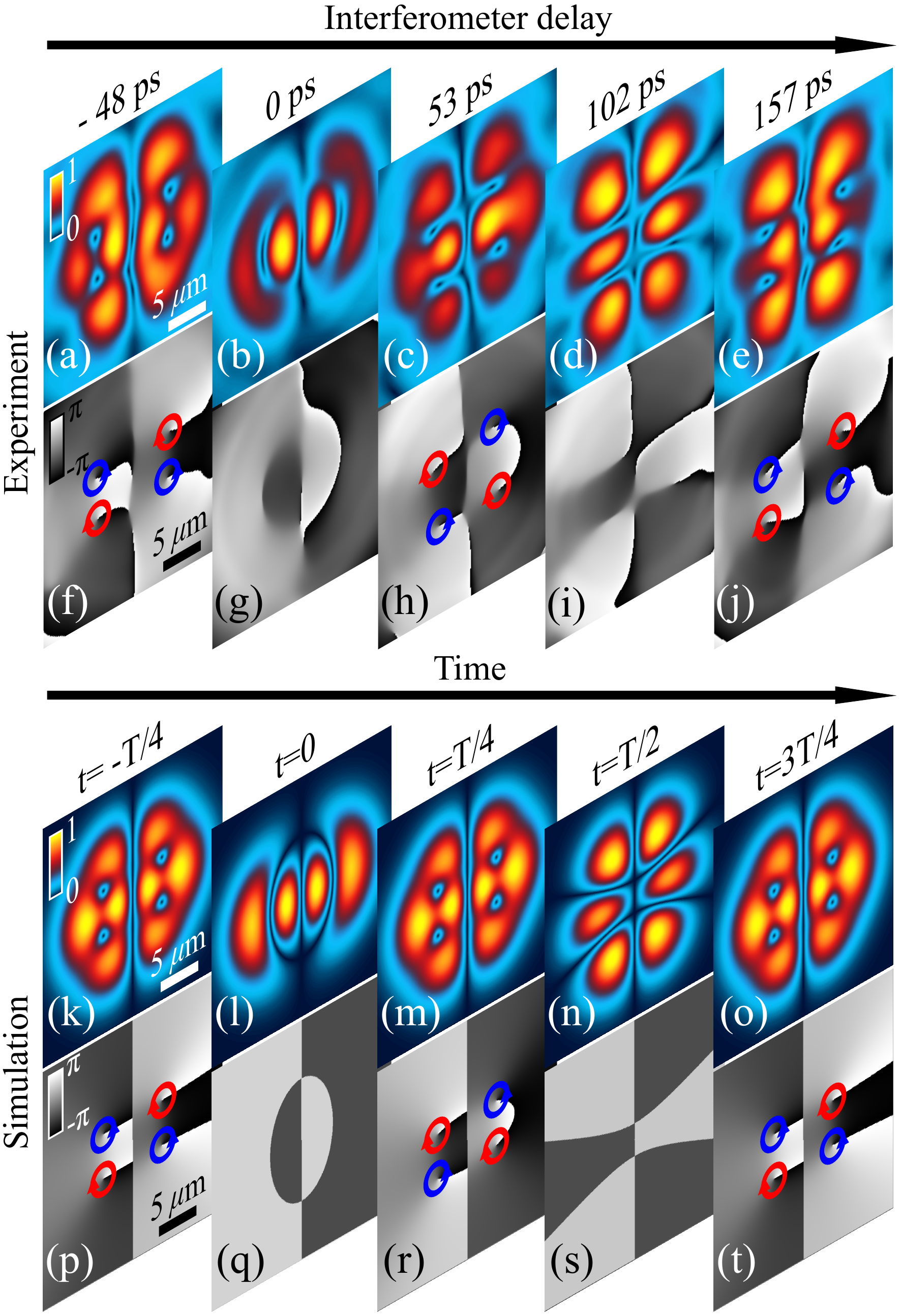}
\caption{\label{fig:epsart} 
Measured normalized temporal evolution of real-space intensity (a-e) and phase (f-j) maps for the trapped condensate. Corresponding modelled evolution of the wavefunction intensity (k-o) and phase (p-t) for a superposition of the complex even Ince-Gaussian functions $\Psi(\mathbf{r},t) = IG_{31}^{(e)}(\mathbf{r})e^{ i \Delta t/2\hbar} +IG_{33}^{(e)}(\mathbf{r}) e^{-i \Delta t/2\hbar} e^{i \phi}$ with an ellipticity parameter $\epsilon$=2. Red and blue circles with arrows in (f,h,j,p,r,t) schematically denote direction of phase winding for generated vortices which periodically flip their topological charges.}
\label{fig3}
\end{figure}

To confirm that the condensate is coherently populating both modes at the same time and undergoing robust dynamical self-oscillations (i.e., nonstationary evolution) we measure the time-resolved spatial first-order coherence function $g^{(1)}(\mathbf{r},\tau)$. For this, at different time delay $\tau$, we record the interference pattern between the total condensate PL with a small uniform spatially expanded region of itself passing through an optical delay line in a modified Mach-Zehnder interferometer configuration. Here, the spatially uniform signal is cut out of the centre of the polariton PL and, when expanded, plays a role of a flat phase reference wave. The reconstructed real-space intensities and phase maps corresponding to different time delay $\tau$ are depicted in Figs.~\ref{fig3}(a-j). We observe periodic flipping of the vortex signs (charges) in the cluster twice per beating period $T \approx 200$ ps. This period corresponds to an energy splitting of $\Delta = 20.7$ $\mu$eV, similar to our estimate based on the homodyne interferometric technique. The observed oscillations evidence that the condensate is indeed occupying two distinct energy modes. Such energy-fractured condensation is typical for nonlinear driven-dissipative systems and has been observed before in nonresonantly generated polariton condensates~\cite{Tosi_NatPhys2012,Lotka2020}. In Supplemental Material, we additionally demonstrate the experimental results for extracted modulus $|g^{(1)}(\mathbf{r},\mathbf{-r},\tau)|$ integrated over the condensate area versus delay time fitted with a simple model describing a weakly interacting two-level bosonic gas. We note that the extracted coherence time of the condensate is $t_c = 426$ ps which is typical for optically trapped condensates~\cite{cohrev}. We note that the excitation beam is circularly polarized and, through additional polarization-resolved measurements (see Supplemental Material), we confirm that the condensate is also strongly circularly polarized which indicates that spin dynamics (e.g., self-sustained Larmor precession) cannot be responsible for the observed oscillations~\cite{cohrev,gnusov}. 
%For comparison, we show in Fig.~\ref{fig3}(a) the measured condensate spectrum from time-integrated measurements (corresponding to Fig.~\ref{fig1}) where the fine mode splitting cannot be fully resolved due to the limited spectral resolution ($\approx$ $40$ $\mu$eV), underlining that such intricate  dynamics in polariton condensates can be easily overlooked.

Figures~\ref{fig3}(k-t) show the calculated temporal evolution of the system described by the nonstationary superposition  $\Psi = IG_{31}^{(e)}e^{ i \Delta t/2\hbar} + IG_{33}^{(e)} e^{-i \Delta t/2\hbar} e^{i \phi}$, where $\Delta = 2 \pi \hbar/T$. One can see that the instantaneous intensity distribution in Fig.~\ref{fig3}(k,m,o) is self-replicating every half oscillation period whereas the phase maps, shown in Fig.~\ref{fig3}(p,r,t), display a flip in the sign of the topological charges. We conclude on a good agreement between experiment and model and believe that such optically engineered polariton system can be used for generation of structured light with arranged phase singularities which periodically evolve.

In addition, we perform generalized Gross-Pitaevskii simulations (mean field treatment) where the polariton condensate wave function $\Psi(\mathbf{r},t)$ is coupled to a driven exciton reservoir $X(\mathbf{r},t)$ rate equation~\cite{Wouters_PRL2007},
\begin{eqnarray}
i\hbar\frac{\partial \Psi}{\partial t}&=&\bigg[(i \Lambda - 1)\frac{\hbar^{2}\nabla^{2}}{2m}
 +\alpha |\Psi|^2 +G\left(X + \frac{\eta P(\mathbf{r})}{\Gamma}\right) \nonumber 
\\
&&  +\frac{i \hbar}{2}\left(R X-\gamma\right)\bigg]\Psi,
\label{gross-pitaevskii_equation}
\\
\frac{\partial X}{\partial t}&=&-\left(\Gamma+R |\Psi|^{2}\right)X +P(\mathbf{r}).
\label{rate_equation_of_reservoir}
\end{eqnarray}
Here, $m$ is the polariton mass, $\gamma^{-1}$ the polariton lifetime, $G = 2 g |\chi|^2$ and $\alpha = g |\chi|^4$ are the polariton-reservoir and polariton-polariton interaction strengths, respectively, $g$ is the exciton-exciton dipole interaction strength, $|\chi|^2$ is the excitonic Hopfield fraction of the polariton, $R$ is the scattering rate of reservoir excitons into the condensate, $\Lambda$ is a phenomenological energy dampening parameter, $\Gamma$ is the reservoir decay rate, $\eta$ quantifies additional blueshift coming from a dark background of excitons which do not scatter into the condensate, and $P(\mathbf{r})$ is the nonresonant continuous-wave pump. We base model parameters on the cavity properties~\cite{doi:10.1063/1.4901814}: $m = 5.64 \times 10^{-5} m_0$ where $m_0$ is the free electron mass, $\gamma^{-1} = 5.5$ ps, $|\chi|^{2}=0.4$ since our cavity is negatively detuned, and a scaled interaction strength $g= 1 \, \mu\mathrm{eV\,\mu m^{2}}$ corresponding to the 6 GaAs-type quantum wells. Remaining parameters are taken similar to those used to describe past experiments, $\hbar R=2.8g$; $\eta=3.6$; $\Lambda = 0.05$; and $\Gamma = \gamma$. The results from simulation are shown in the Supplementary Animation displaying a limit cycle behaviour in the condensate dynamics, corresponding to the evolution in Fig.~\ref{fig3}(k-t). Here, we use a slightly elliptical trap profile $P(\mathbf{r}) =P_0 L^4 /[(r^2 - r_0^2)^2 + L^4]$ to emulate the energy-split mode structure with pump parameters: $r^2 = (x/a)^2 + (y/b)^2$; $a/b = 1.15$; $r_0 = 10$ $\mu$m; $L = 7$ $\mu$m; $P_0 = 22$ ps$^{-1}$ $\mu$m$^{2}$.

In this work, we provided conclusive evidence of spontaneous vortex cluster formation in a nonresonantly driven polariton condensate. Our findings are accurately reproduced through linear superposition of the trap high-order Ince-Gaussian modes. We also find, through two independent techniques, that the condensate is energy-fractured across two split trap modes which leads to rapid density self-oscillations and periodic sign flipping of the topological charges in the generated vortex cluster. We stress that observed arranged vortices do not appear from direct phase imprinting of angular momentum like in Refs.~\cite{Sanvitto_NatPhys2010, Donati_PNAS2016, Boulier2016, Dominici_NatComm2018, Panico2021}, or any trap rotation like in atomic Bose-Einstein condensates~\cite{Fetter_RMP2009}, but rather spontaneously form and freely evolve. Our work opens exciting perspectives on designing complex structured light sources with periodically evolving singular phase patterns in the strong light-matter coupling regime, beyond the higher-order modes considered here. 

The datasets presented in this paper are openly
available from the University of Southampton repository~\cite{data}.

The reported study was funded by RFBR according to the research project No. 20-32-90128.
The authors acknowledge the support of the UK’s Engineering and Physical Sciences Research Council (grant EP/M025330/1 on Hybrid Polaritonics), and European Union’s Horizon 2020 program, through a FET Open research and innovation action under the grant agreement No. 899141 (PoLLoC). H.S. acknowledges the Icelandic Research Fund (Rannis), grant No. 217631-051.

\providecommand{\noopsort}[1]{}\providecommand{\singleletter}[1]{#1}%

\end{document}